\documentclass[a4paper,aip,rsi,reprint,graphicx,hyperref]{revtex4-1} 
\usepackage{amsmath,amssymb,graphicx}

\def \sourcedist {9.77 cm}

\begin{document}

\preprint{LA-UR-14-24574}

\title{Radial distribution of charged particles in a magnetic field} 

\author{S.~K.~L. Sjue}
\email[]{sjue@lanl.gov}
\author{L. Broussard}
\author{M. Makela}
\author{P.~L. McGaughey}
\affiliation{Los Alamos National Laboratory}
\author{A.~R. Young}
\affiliation{North Carolina State University}
\author{B.~A. Zeck}
\affiliation{Los Alamos National Laboratory}
\affiliation{North Carolina State University}

\date{\today}

\begin{abstract}
The radial spread of charged particles emitted from a point source in a magnetic field is a potential source of systematic error for any experiment where magnetic fields guide charged particles to detectors with finite size.  Assuming uniform probability as a function of the phase along the particle's helical trajectory, an analytic solution for the radial probability distribution function follows which applies to experiments in which particles are generated throughout a volume that spans a sufficient length along the axis of a homogeneous magnetic field.  This approach leads to the same result as a different derivation given by Dubbers \emph{et al}.\cite{DubbersNIM}  But the constant phase approximation does not strictly apply to finite source volumes or fixed positions, which lead to local maxima in the radial distribution of emitted particles at the plane of the detector.  A simple method is given to calculate such distributions, then the effect is demonstrated with data from a $^{207}$Bi electron-conversion source in the superconducting solenoid magnet spectrometer of the Ultracold Neutron facility at the Los Alamos Neutron Science Center.  Potential future applications of this effect are discussed.
\end{abstract}

\pacs{23,29,41}

\maketitle 

\section{Experimental motivation}

Magnetic spectrometers for charged particles are widely used in nuclear and particle physics experiments.  PERKEO \cite{PERKEO,PERKEOII,PERKEOIII} and UCNA \cite{PlasterUCNA,scsNIM} are two examples: these experiments use magnetic spectrometers to measure the beta asymmetry associated with the decay of polarized neutrons, which characterizes the correlation between the spin of the decaying neutron and the momentum of the emitted electron,
\begin{equation}
\dfrac{dN_e}{d(\cos\theta_e)}=N_0(E_e) 
\left(1+A\dfrac{p_e}{E_e}\cos\theta_e \right),
\end{equation}
where $\theta_e$ is the angle between the neutron's spin and the electron's momentum and ``A'' is known as the beta asymmetry coefficient.  Another current example is KATRIN,\cite{KATRIN} which aims to find (or determine an upper limit on) the mass of the electron neutrino by measuring the most energetic electrons from the beta decay of tritium.

In each of these experiments, beta decay takes place in some volume in a magnetic field, then the decay electrons are guided along the magnetic field lines to a detector at one end of the decay volume.   The PERKEO and UCNA experiments effectively cover 100\% of 4$\pi$ steradians in solid angle for neutrons that decay within the fiducial volume with detectors at both ends of the decay volume.  However, for a decay event on a magnetic field line that passes near the edge of a detector, it is possible for the electron to miss the detector.  The probability for such a missed event depends upon the radial probability distribution of these particles, which is a function of the particle's charge, momentum, pitch angle and positon along the magnetic field line with respect to the detector.  A convolution between this distribution and the experimental arrangement (neutron source density, magnetic field and detector geometry) determines the size of this potential systematic error for these measurements.

The radial distribution of particles in question exhibits some peculiar features, particularly for sources at a fixed distance from a detector.  The fixed distance radial distributions are not monotonic and they exhibit multiple local maxima.  The following sections elucidate these features then present an experimental demonstration of the effect.

\section{Equations of motion}

\begin{figure}
\include{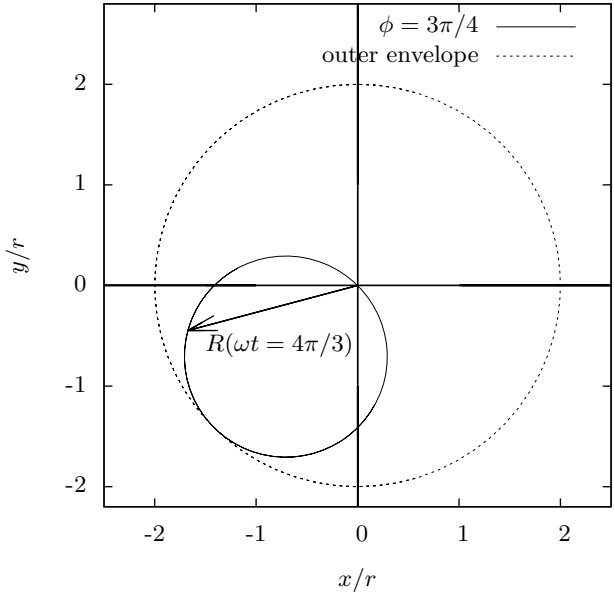}
\caption{Trajectory from Equation \ref{eq:rPerp} for $\phi=3\pi/4$, along with the outer envelope of all possible trajectories including the complete range in $\phi$, $[0,2\pi]$.  The arrow shows the position for this trajectory at a time when $\omega t = 3\pi/4$.}
\label{fig:trajectory}
\end{figure}

The radius of a charged particle in a magnetic field with momentum perpendicular to the field is
\begin{equation}
r_0=\dfrac{p}{qB}.
\label{eq:bRad}
\end{equation}
Units in which the speed of light is unity ($c=1$) are used here and throughout.  If the momentum vector makes an angle $\theta$ with the field, then the radius is
\begin{equation}
r=r_0\sin\theta.
\label{eq:rOfTheta}
\end{equation}
The $\hat z$ axis is defined to follow the magnetic field, so $\theta$ as defined is the polar angle and $\phi$ will be the azimuthal angle.  In a homogeneous magnetic field, the position of the charged particle as a function of momentum, direction and time is
\begin{align}
\vec r(t) = &\hat x [r(1-\cos\omega t)\cos\phi+r\sin\omega t \sin \phi ]\nonumber\\
+&\hat y[-r(1-\cos\omega t)\sin\phi+r\sin\omega t \cos \phi]\\
+&\hat z (p/E) t \cos\theta, \nonumber
\label{eq:r3d}
\end{align}
in which the angular frequency is given by $\omega=qB/m$ and it has been assumed that the particle starts at $z=0$.  This equation describes positively chaged particles, but it is only necessary to change the sign on either the $x$ or $y$ component to describe negatively charged particles.  One such trajectory in the $x-y$ plane for $\phi=3\pi/4$ is shown in Figure \ref{fig:trajectory}.  The magnitude of the radial position in the plane perpendicular to the magnetic field as a function of time is independent of the azimuthal angle.  The quantity $\sqrt{\vec x \cdot \vec x + \vec y \cdot \vec y}$ gives the magnitude of $\vec R$ as a function of time:
\begin{equation}
R(t)=r\sqrt{2-2\cos \omega t}.
\label{eq:rPerp}
\end{equation}
This expression could also be written as the absolute value of a sine function.  This equation is not sensitive to the sign of a particle's charge.

\section{Phase averaged radial distributions for monoenergetic particles}

For the sake of insight, a simple approximation will be discussed first.  This approximation confirms results obtained from a different approach by Dubbers,\cite{DubbersNIM} while its simplicity illuminates the nature of the approximations that were made.  If all values of the phase are equally likely, then the probability for finding the particle at a particular value of $R$ must be proportional to the time spent there during any given revolution in the helical trajectory.  In other words, $P(R)\propto dt/dR$.  The derivative of Equation \ref{eq:rPerp} with respect to $t$ can be inverted and simplified to find
\begin{equation}
\dfrac{dt}{dR}=
\dfrac{1}{\omega r}\dfrac{1}{\sqrt{1-(R/2r)^2}}.
\label{eq:dtdR}
\end{equation}
The denominator in front of the radical is just the particle's velocity in the $x-y$ plane.  All values of $R$ are covered by one half of the smaller circle shown in Figure \ref{fig:trajectory}, which requires a time $\Delta t = \pi/\omega$.  The normalized distribution is given by $(1/\Delta t)dt/dR$:
\begin{equation}
\dfrac{dP}{dR}=
\dfrac{1}{\pi r}\dfrac{1}{\sqrt{1-(R/2r)^2}}.
\label{eq:dPdR}
\end{equation}
Inspection of Figure \ref{fig:trajectory} corroborates this solution: the probability is smallest at $R=0$, where the velocity is parallel to $\vec R$; the probability diverges at $R=2r$, where the particle's velocity is actually perpendicular to $\vec R$ and $R$ is instantaneously constant.

\begin{figure}
\include{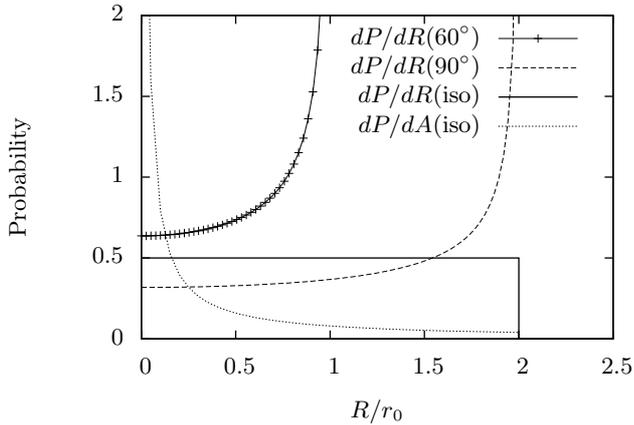}
\caption{Radial probability distributions averaged over the phase.  The distributions with the label ``iso'' have been averaged over the solid angle of an isotropic distribution.  All distributions are zero for $R/r_0 > 2$.}
\label{fig:padists}
\end{figure}

Equation \ref{eq:dPdR} implicitly includes the pitch angle $\theta$ in the variable $r$.  Written explicitly as a function of these variables, the radial distribution for a given angle is 
\begin{equation}
\dfrac{dP}{dR}(R,\theta)=
\dfrac{1}{\pi r_0}\dfrac{1}{\sqrt{\sin^2\theta-(R/2r_0)^2}}.
\label{eq:dPdRofTheta}
\end{equation}
Up to a normalization factor, this result and its integrals agree with Equations 20 and 25 of Dubbers.\cite{DubbersNIM}  So it is safe to conclude that the approximations made to arrive at that result are equivalent to assuming uniform probability in the phase along the particle's trajectory.  The radial distribution for isotropically emitted monoenergetic particles can be found by averaging this expression over the solid angle, with the limits of integration for the polar angle set to $\theta_<= \arcsin(R/2r_0)$ (the minimum angle that can reach $R$) and $\theta_> =\pi/2$.  The integral takes a simple form upon substitution of $u=\cos\theta/\sqrt{1-(R/2r_0)^2}$, which gives the radial distribution of isotropically emitted monoenergetic particles,
\begin{align}
\left(\dfrac{dP}{dR}\right)_\textrm{isotropic}=
\dfrac{1}{\pi r_0}\int_0^1 \dfrac{du}{\sqrt{1-u^2}}=
\dfrac{1}{2r_0}.
\label{eq:dPdRiso}
\end{align}
Hence all values of $R$ from 0 to $2r_0$ are equally likely for a monoenergetic, isotropic source of particles.  The density of particles per unit area from such a source follows from multiplying by $dR/dA$, from which one finds\begin{equation}
\left(\dfrac{dP}{dA}\right)_\textrm{isotropic}=\dfrac{1}{4\pi R r_0},
\end{equation}
which implies that the density falls as $1/R$.  The distributions derived in this section are shown in Figure \ref{fig:padists}.

\section{Radial distributions at a fixed distance}

\begin{figure}
\include{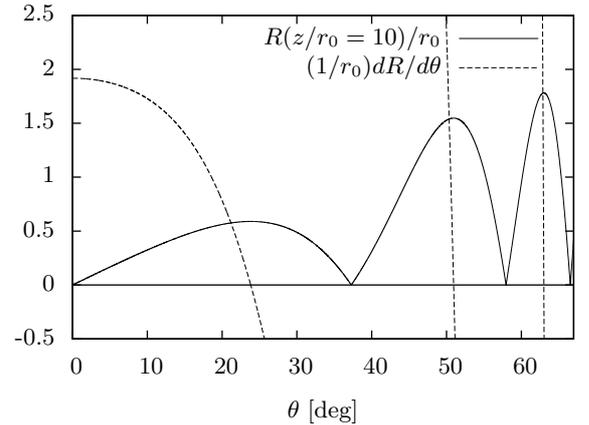}
\caption{$R$ and $dR/d\theta$ for $z/r_0=10$.  Neither $R$ nor its derivative are one-to-one functions.  A more sophisticated procedure is necessary to calculate the probability as a function of $R$ in this case.}
\label{fig:z10fs}
\end{figure}

The phase averaged results are not applicable for a fixed distance between the source of particles and the detector.  The total phase angle at distance $z$ from the origin can be written in terms of the pitch angle,
\begin{equation}
\omega t = \dfrac{z}{r_0\cos\theta},
\end{equation}
in order to express the quantity of interest $R$ from Equation \ref{eq:rPerp} explicitly as a function of the distance and the pitch angle:
\begin{equation}
R(z,\theta)=r_0\sin\theta\sqrt{2-2\cos\left(\dfrac{z}{r_0\cos\theta}\right)}.
\label{eq:rPerpFixedZ}
\end{equation}
There can be no distribution analogous to Equation \ref{eq:dPdRofTheta}, because $R(z,\theta)$ is completely determined by $z$ and $\theta$.   A radial distribution as a function of pitch angle only makes sense if the particles come from a range of positions within the magnetic field.  One might be tempted to take the derivative of $R(z,\theta)$ with respect to $\theta$, invert it, then average it over the solid angle to find an isotropic distribution analogous to Equation \ref{eq:dPdRiso}.  Such an average to obtain an isotropic distribution is not possible without partitioning the integral, because $dR/d\theta$ is not a one-to-one function within the limits of integration.  Both $R$ and $dR/d\theta$ are shown in Figure \ref{fig:z10fs}.

The radial distribution is more difficult to calculate for a fixed source-to-detector distance, but several useful observations can be made about its form.  In the limit $\theta\rightarrow\pi/2$, the phase $(z/r_0)/\cos\theta$ oscillates rapidly and the phase-averaged limit of Equation \ref{eq:dPdRiso} will be recovered.  The range of angles over which this limit is approached depends upon the ratio $z/r_0$; in the limit $z/r_0\rightarrow\infty$, the phase-averaged limit will become valid over the whole range.   The largest departure from the phase-averaged limit comes at small angles.  The zeroes of $R(z,\theta)$ come at $\theta=0$ and at $\cos\theta=z/(2r_0n\pi)$, with $n\in\mathbb{Z}$ satisfying $\cos\theta<1$.  The maxima in the radial distribution are found between the zeroes in $R(z,\theta)$ (corresponding to the maxima of $R(z,\theta)$) and their amplitude is proportional to the distance between the zeroes.  The derivative at $\theta=0$ is:
\begin{equation}
\dfrac{1}{r_0}\left|\dfrac{dR}{d\theta}\right|=
\left|\sin\left(\dfrac{2z}{r_0}\right)\right|.
\end{equation}
The distribution will have its largest maximum at a larger radius if $2z/r_0$ is an integer multiple of $\pi$ and a smaller radius if $2z/r_0$ is a half-integer multiple of $\pi$.

\begin{figure}
\include{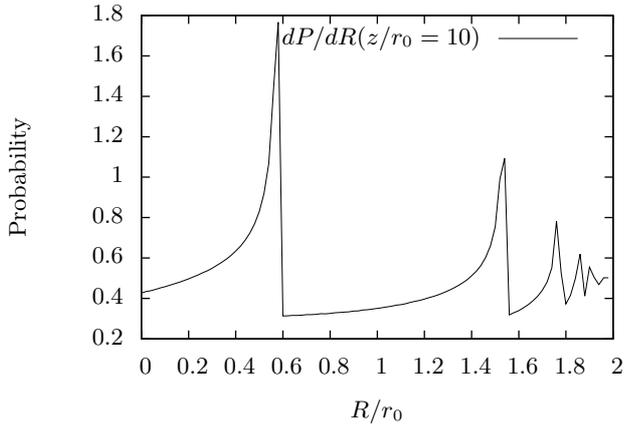}
\caption{$dP/dR$ for $z/r_0=10$ from Equation \ref{eq:numericalIntegration}.  The largest maximum in the distribution comes from the broadest maximum in $R$ as a function of angle shown in Figure \ref{fig:z10fs}.  The distribution approaches the phase-averaged constant limit in the upper limit of the range of $R$.}
\label{fig:z10prob}
\end{figure}

An analytic solution to $\int d(\cos\theta$)/dR has not been found and even if one were known, it would be necessary to partition it and integrate separately between each sequential pair of zeroes in $R(z,\theta)$.  Despite this complication, it is simple to calculate the distribution by performing the integral numerically or with a Monte Carlo study.  This procedure is described by an integral equation:
\begin{equation}
\dfrac{dP}{dR}(R,z)=
\left[ \int_0^1 d(\cos\theta) \right]_{R(\theta,z)\in[R,R+dR]}
\label{eq:numericalIntegration}
\end{equation}
Either random or uniform values of $\cos\theta$ can be used to calculate a value of $R$ from Equation \ref{eq:rPerpFixedZ}.  Then the corresponding histogram in the range $[R,R+dR]$ is incremented.  The probability as a function of radius for $z/r_0=10$ is shown in Figure \ref{fig:z10prob}.

\section{Experimental demonstration}

In order to verify these sharply peaked distributions,  measurements were made using the apparatus built for the UCNB experiment in the superconducting solenoid (SCS) magnetic spectrometer\cite{scsNIM} at the Los Alamos National Laboratory Ultracold Neutron source.\cite{UCNsource1,UCNsource2}  A $^{207}$Bi source was attached to the end of the high voltage shroud, centered in front of a Si detector with 127 pixels\cite{americo} at a distance of \sourcedist.  The current in the magnet was incremented from full field to less than 25\% of the full field.  Spectra were taken from the center pixel and in pixels from the first two rings around the center for each setting of the magnetic field.  Looking at monoenergetic conversion electrons from the source, these spectra amount to experimentally varying $z/r_0$ while measuring the integral of $d(\cos\theta)/dR$ over the area ($\approx 0.78$ cm$^2$) of the hexagonal pixels on the silicon detector.  The data shown here correspond to values of $z/r_0$ ranging from 3.1 to 12.5; $B=0.46$ T results in the functions shown in Figures \ref{fig:z10fs} and \ref{fig:z10prob}.  The source and detector were placed such that the electrons travelled through a primarily homogeneous field between source and detector, according to the specifications of the SCS and its field map.  For this $\sim 10$ cm distance, the maximum specified field would correspond to a uniform field of 0.6 T for the first 9 cm, then a decrease in the field from 0.6 T to 0.59 T over the last cm.

\begin{figure}
\include{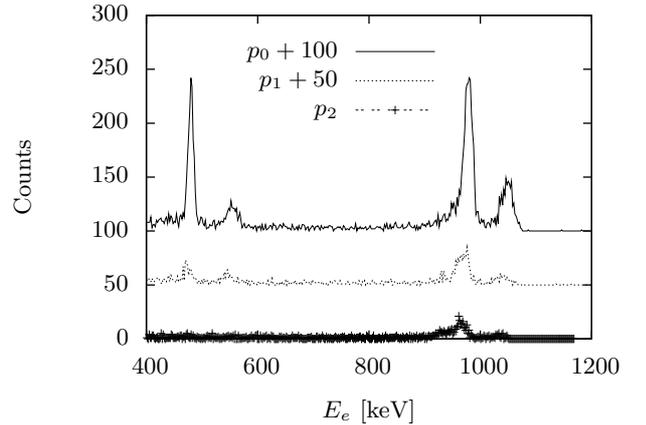}
\caption{Spectra from three pixels with the magnet's current set to 65 A (corresponding to 0.315 T).  The three pixels $p_0$, $p_1$ and $p_2$ are respectively at $r=0$, 1 and 2 cm from the center of the detector.  Note that for the two conversion electrons at $E_e\approx 482$ and 554 keV, there are no counts in $p_2$ because it is past $2r_0$ for these energies at this magnetic field.}
\label{fig:bi207}
\end{figure}

Spectra from three channels at one magnetic field setting are shown in Figure \ref{fig:bi207}.  The quantity of interest is the number of counts in the conversion electron peak at roughly 976 keV.  This peak is used because it has 7.1\% intensity and it suffers minimally from backscattering and background noise due to its relatively high energy.  

For each step in the magnetic field, the data were taken for five minutes on all instrumented pixels.  The number of counts in each pixel ($p_i$) is given by a modified version of Equation \ref{eq:numericalIntegration},
\begin{equation}
N_i= \dfrac{dN}{dt} \Delta t\left[ \int d\Omega \right]_{\vec R(\phi,\theta,z)\in \{p_i\}},
\end{equation}
where $dN/dt$ is the decay rate, $\Delta t$ is the duration of the run, $\theta$ is still the pitch angle and $\phi$ is the azimuthal angle.  Uncertainty related to the length of time for a run is not a factor when considering the ratio of counts between two pixels.

\begin{figure}
\include{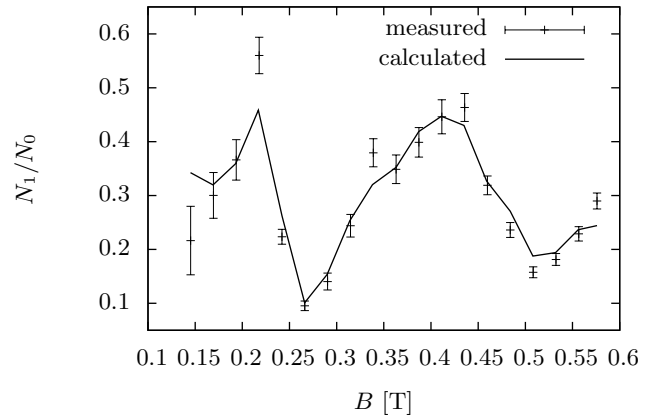}
\caption{Ratio between counts in the 976 keV peak from $p_1$ and $p_0$ in Figure \ref{fig:bi207} as a function of magnetic field.}
\label{fig:ratio1}
\end{figure}

\begin{figure}
\include{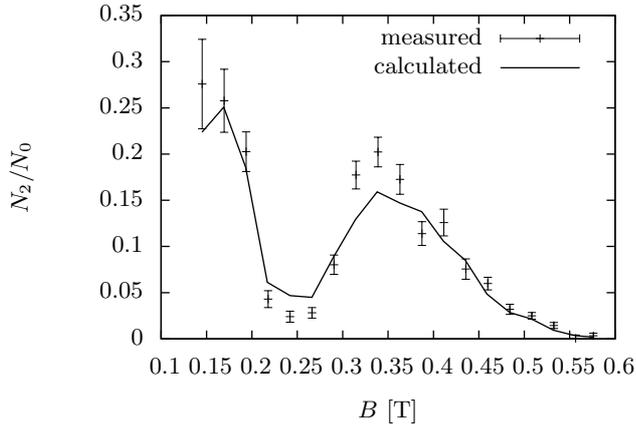}
\caption{Ratio between counts in the 976 keV peak from $p_2$ and $p_0$ in Figure \ref{fig:bi207} as a function of magnetic field.}
\label{fig:ratio2}
\end{figure}

Results from the field scan are shown in Figures \ref{fig:ratio1} and \ref{fig:ratio2}.  Only the position of the source, which was attached to the experiment with a piece of aluminized mylar tape, was allowed to vary by a small amount while making calculations to compare with the data.  The coordinates in best agreement with the data from those sampled were for the source displaced 1.1 mm from the center of the detector.  A very small correction was calculated due to a slight decrease in the magnetic field from 0.576 T by about 0.01 T over the last centimeter to the detector (or an equivalent fractional amount for smaller field settings), but this correction has a negligible effect on the quality of the results.  No correction has been applied for backscattering of electrons from the detector.

\section{Conclusions}

The calculated distributions capture the trends in the data quite well.  Apparently, the radial distribution of isotropically emitted charged particles in a magnetic field does exhibit sharp maxima depending on the distance between the source and the detector.  Such maxima could also have an effect on an asymmetry measurement, given a small decay volume or a decay volume with regions where $r_0/z$ is not sufficiently large.  Analyses of experiments intended to measure angular distributions of charged particles in magnetic fields with detectors of finite size should consider this effect.  The strategy employed by the UCNA collaboration has been to vary the fiducial volume to verify that the effect on the measured beta asymmetry is negligible, which is possible because of the position sensitivity of the multi-wire proportional chambers\cite{mwpc} before the scintillator detectors.

Backscattering of electrons and protons from silicon detectors is a significant source of systematic error for precise experimental measurements of beta decay parameters.  This is true for several current experiments, including decay-in-flight measurements of the neutron lifetime\cite{NISTlifetime} and the electron-antineutrino correlation.\cite{aCORN}  Electron backscattering from silicon detectors is a potential source of systematic errors for KATRIN as well.  The corrections due to backscattering are generally made based on simulation packages such as PENELOPE\cite{PENELOPE} and GEANT.\cite{GEANT4}  The backscattering of electrons from silicon detectors has only been measured at normal incidence for a limited range of energies.\cite{backscatter1,backscatter2}  

It would be worthwhile to verify the angular dependence of backscattering from these simulations, which is quite large.  For example, PENELOPE simulations indicate that the 976-keV conversion electrons presented in this study have a probability of backscattering from bulk silicon that varies from 13\% at normal incidence to more than 90\% at grazing angles.  The effect on the data presented in this study does not appear to be as large as these probabilities would suggest, probably due to a magnetic mirror effect: the full field in the magnetic spectrometer at the source and detector is about 0.6 T, compared to 1 T in the center of the spectrometer.  A large fraction of the backscattered electrons should reflect from the pinch in the magnetic field then onto the detector again.  The experimental technique used to take the data presented in this paper, together with ~10 nanosecond timing resolution to resolve backscattering on one pixel or from one pixel to another, might be used to study the probability of electron backscattering from bulk silicon as a function of incidence angle.  Another possibility is a slightly modified apparatus to absorb electrons before they can reflect from the magnetic mirror.  Characterization of such effects is crucial to achieve the $10^{-4}$ resolution necessary for maximum sensitivity to possible new physics effects on neutron decay parameters.

\section*{Acknowledgement}

We would like to express our gratitude to Z. Wang of Los Alamos National Laboratory for early development of the data acquisition system used to take the data presented here, along with some helpful feedback.

\end{document}